%To: hep-th@xxx.lanl.gov
%Subject: put
%
%\\
%DISPERSION RELATIONS IN STRING THEORY, Eric D'Hoker and D.H. Phong, 17 pages,
%COLUMBIA-YITP-UCLA/93/TEP/45
%\\
%We analyze the analytic continuation of the formally divergent one-loop
%amplitude for scattering of the graviton multiplet in the Type II Superstring.
%In particular we obtain explicit double and single dispersion relations,
%formulas for all the successive branch cuts extending out to $+\infty$,
%as well as for the decay rate of a massive string state of arbitrary mass $2N$
%into two string states of lower mass. We compare our results with
%the box diagram in a superposition of $\phi^3$-like field theories.
%The stringy effects are traced to a convergence problem in this superposition.
%\\
%
%%%%%%%%%%%%%%%%%%%%%%%%%%%%%%%%%%%%%%%%%%%%%%%%%%%%%%%%%%%%%%%%%%%%%%%%%%%%%%%
%%      There is a postscript file attached to the end of the file, which    %%
%%      has been compressed and uuencoded with 'uufiles'. Save everything    %%
%%      after the line '%%%DETACH HERE' into a file symm.uu, and follow      %%
%%      the instructions there.                                              %%
%%      You need the macro epsf.tex to generate the figure.                  %%
%%      This is a TeX file and requires the macro PHYZZX.                    %%
%%%%%%%%%%%%%%%%%%%%%%%%%%%%%%%%%%%%%%%%%%%%%%%%%%%%%%%%%%%%%%%%%%%%%%%%%%%%%%%
%
%\magnification=1200
\baselineskip=10pt
\overfullrule=0pt

\input epsf
\def\Re{{\rm Re}}
\def\Im{{\rm Im}}
\input phyzzx

\rightline{COLUMBIA-YITP-UCLA/93/TEP/45}
%\rightline{hep-th/9404128}

\bigskip
\bigskip

\centerline{{\bf DISPERSION RELATIONS IN STRING THEORY}
\footnote{*}{ Research Supported in part by NSF grants DMS-92-04196
and PHY-92-18990.}}
\bigskip
\bigskip
\centerline{{\bf Eric D'Hoker}
\footnote{**} {Electronic Mail Address: DHOKER@UCLAHEP.BITNET}}

\centerline{{\it Yukawa Institute for Theoretical Physics}}
\centerline{{\it Kyoto University}}
\centerline{{\it Kyoto 606, JAPAN}}

\medskip
\centerline{{and}}
\medskip

\centerline{{\it Physics Department}}
\centerline{{\it University of California, Los Angeles}}
\centerline{{\it Los Angeles, California 90024-1547, USA}}

\bigskip

\centerline{{\bf D. H. Phong}\footnote{***}
{Electronic Mail Address: PHONG@MATH.COLUMBIA.EDU}}
\centerline{{\it Mathematics Department}}
\centerline{{\it Columbia University}}
\centerline{{\it New York, N.Y. 10027, USA}}

\bigskip

\centerline{\bf ABSTRACT}
\bigskip
We analyze the analytic continuation of the formally divergent one-loop
amplitude for scattering of the graviton multiplet in the Type II Superstring.
In particular we obtain explicit double and single dispersion relations,
formulas for all the successive branch cuts extending out to $+\infty$,
as well as for the decay rate of a massive string state of arbitrary mass $2N$
into two string states of lower mass. We compare our results with
the box diagram in a superposition of $\phi^3$-like field theories.
The stringy effects are traced to a convergence problem in this superposition.
\vfill\eject

\centerline{\bf 1. Introduction}
\bigskip
A fundamental direction in the exploration of superstring theory is the
understanding of its perturbative expansion in the number of string loops.
Already substantial progress was made in obtaining a consistent formulation
of perturbation theory as a sum over Riemann surfaces. In particular,
it was shown that amplitudes defined this way are Lorentz invariant and
perturbatively unitary [1], and it is generally believed that order by
order, superstring loop amplitudes are finite. In particular, it was argued
long ago that the Type II and heterotic amplitudes
do not exhibit the tachyon and massless dilaton tadpole
divergences that are known to occur in the bosonic string [2][3][4].
\medskip
Our understanding is, however, still rather incomplete compared to
the situation in quantum field theory. There, simple scaling arguments
and recursive combinatorics guarantee a simple physical picture of
renormalizability
in perturbation theory. Feynman's $i\epsilon$ prescription on the propagators,
combined with Cutkovsky and Landau cutting rules, provides a simple picture
of unitarity and causality. In superstring theory, the
rules of perturbation theory do not exhibit the properties of renormalizability
(or finiteness) and unitarity (or causality) manifestly. Instead unitarity of
the
amplitudes was established
only indirectly, by showing equivalence between the covariant and the
light-cone
formulations.
Renormalizability or finiteness have not even reached such a level of indirect
understanding.
\medskip
In an earlier work [5], we had shown that the formally real and divergent
one-loop
amplitude for the scattering of four external states in the graviton multiplet
of the Type II
superstring can be analytically continued to a finite, complex, amplitude
consistent with the optical theorem. We also gave a simple explicit formula
for the leading term in the forward
scattering amplitude. Here we extend this work further by producing explicit
formulas for all the successive branch cuts extending out to $+\infty$ as well
as for the decay rate of a massive string state of arbitrary mass $2N$ into
two string states of lower mass. We illustrate our
results by comparing them with those in an infinite superposition of
$\phi^3$-like field theories.
The stringy effects are traced to a convergence problem in this superposition.
The key ingredients in our approach are dispersion relations, which occur in
the form of both double and single relations. Our methods apply in more general
set-ups, including the heterotic string and certain toroidal, orbifold,
and Calabi-Yau compactifications.
We shall however present these extensions and other applications elsewhere.
\bigskip
\centerline {\bf 2. The Analytic
Continuation Problem for Superstring Scattering Amplitudes}
\bigskip
We begin by discussing the Type II superstring amplitudes
for the scattering of four massless on-shell states in the graviton multiplet,
their domain
of convergence, and the physical meaning of their analytic continuation.
They are of the form
$$
\epsfysize1.3in\vcenter{\epsfbox{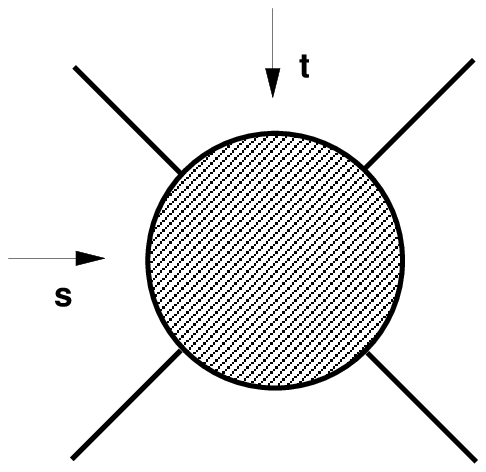}}
= g^4 \delta ^{(10)} (\sum _{i=1}^4 k_i)
\epsilon_1\epsilon_2\epsilon_3\epsilon_4K
\bar K\,
A(s,t,u)\eqno(2.1)
$$
Here $k_i$, $\epsilon_i$, $i=1,\cdots,4$ are respectively the momenta and
polarization
tensors of the incoming particles, $g$ is the string coupling, and $K\bar K$
is a purely kinematical factor analytic
(actually polynomial) in the momenta. The expression $A(s,t,u)$ is the reduced
Lorentz invariant
amplitude, and $s$, $t$, and $u$ are the Mandestam variables defined by
$$
s=s_{12}=s_{34}, \ t=s_{23}=s_{14}, \ u=s_{31}=s_{24}
$$
with $s_{ij}\equiv -(k_i+k_j)^2$. Momentum conservation implies that
$$
s+t+u=0\eqno(2.2)
$$
At tree level, the reduced amplitude is given by the well-known expression in
terms of Green's functions on the complex plane
$$
A_0(s,t,u)={1\over t^2}\int_{\bf C}d^2z|z|^{-s-2}|1-z|^{-t}\eqno(2.3)
$$
The above integral converges only when
$$
\Re\, s<0,\ \Re\, t<2,\ \Re\, (s+t)>0
$$
so that even this case requires an analytic continuation.
It is however readily available since (2.3) can be rewritten in terms of
$\Gamma$
functions as
$$
A_0(s,t,u) ={\Gamma(-{s\over 2})\Gamma(-{t\over2})\Gamma(-{u\over 2})
\over\Gamma(1+{s\over 2})\Gamma(1+{t\over 2})\Gamma(1+{u\over 2})}\eqno(2.4)
$$
Thus the amplitude can be extended as a meromorphic function of $s$ and $t$
throughout the complex plane, with as only singularities simple poles in $s$,
$t$,
and $u=-(s+t)$ at positive even integers. We observe that the integral
representation
for the amplitude produces only a holomorphic function in its region of
convergence,
and that physically important singularities such as poles (corresponding to
massive intermediate states going on mass shell) must arise through analytic
continuation.

\bigskip
To one-loop order, the reduced
amplitude is given by an integral over the moduli space of tori
$$
A_1(s,t,u)=\int_F{d^2\tau\over\tau_2^2}
\int_{M_{\tau}}{d^2z_i\over\tau_2}\prod_{i<j}exp(s_{ij}G(z_i,z_j)/2)\eqno(2.5)
$$
Here $M_{\tau}$ is the torus represented in the plane by the parallelogram with
corners $0$, $1$, $\tau$, and $1+\tau$
and opposite
sides identified, $F$ is the fundamental domain
for the modulus $\tau$
$$F=\{\tau\in{\bf C};\tau_2>0,|\tau_1|\leq1/2,|\tau|\geq1\}$$
and $G(z,w)$ is the scalar Green's function on $M_{\tau}$. It can be expressed
in terms of
Jacobi $\vartheta$-functions as
$$
G(z,w)=-{\rm ln}\big|{\vartheta_1(z-w|\tau)\over\vartheta_1'(0|\tau)}
\big|^2+{2\pi\over\tau_2}
(\Im(z-w))^2
$$
It is easy to see that the integral in (2.5) converges exactly when
$$
\Re\, s=\Re\, t=\Re\, u=0\eqno(2.6)
$$
in which case the exponentials in (2.5) are all phases. In fact (2.5) exhibits
two types of singularities. When $z_i\sim z_j$,
$G(z_i,z_j)\sim-{\rm ln}|z_i-z_j|^2\rightarrow + \infty$, and integrability in
$z_i$ requires $\Re\, s_{ij}<2$.
On the other hand, when $z_i$ and $z_j$ are well-separated, and $\tau_2
\rightarrow \infty$,
$G(z_i,z_j)$ tends to $-\infty$, and integrability in $\tau_2$ requires
$\Re\, s_{ij}\geq0$.
Our assertion follows from the fact that these conditions must hold for all
$i,j$ and that $s+t+u=0$. Thus the amplitude in both the deep Euclidian region
$s<0$ and the physical region $s>0$ must be obtained by analytic continuation.

To appreciate the physical meaning of such analytic continuations, we compare
the
situation
with quantum field theory. There, causality is equivalent to
locality and, combined with Lorentz invariance, implies analyticity of the
Green's
functions in momentum space. The standard way to establish this is to use
locality
and the fact that observable fields commute at space-like separations to derive
a spectral representation for
the Green's functions. The simplest one is the K$\ddot{\rm a}$llen-Lehmann
representation for the two-point function, say for a scalar field,
$$
\epsfysize.75in\vcenter{\epsfbox{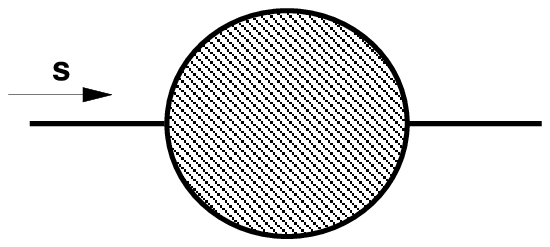}}
 = \int_0^{\infty}dM^2{\rho(M^2)\over s-M^2+i\epsilon}
$$
Note that the integration is over $M^2\geq0$, as the existence of tachyonic
states would automatically violate causality.
Provided this spectral representation is convergent, it immediately implies
that the two-point function is an analytic function of $s$, with only
singularities along the positive
real axis. These singularities may be poles or branch cuts, resulting
respectively from one-particle or multiparticle intermediate
states. In analytic S-matrix theory, the analyticity is assumed as one of the
starting points of the theory [6][7].
\medskip
On the other hand, string theory does not have at present a
formulation in terms of vacuum expectation values of observables. It is not
known whether spectral representations analogous to those of quantum field
theory can be derived solely from
the requirements of causality (i.e. locality of the string interactions) and
Lorentz
invariance. Thus the analyticity of string amplitudes must be established.
Our goal
is to do so, starting from the integral representations (2.5) obtained via
string perturbation theory. More precisely, we address the following questions:
\medskip
\item{(a)} Does an analytic continuation exist?
\item{(b)} Is it unique?
\item{(c)} Is the analytic continuation physically acceptable?
\medskip
We have answered (a) in the affirmative by constructing explicitly
a suitable analytic continuation to the cut plane $s,t,u\in{\bf C}
\setminus{\bf R}_+$, $s+t+u=0$ [5]. In particular, there are no lacunae in the
domain of holomorphy. The answer to (b) is evidently yes, since in each
variable
(say $s$), two analytic continuations would have to agree on a line in the
complex
plane (imaginary $s$), and thus must be equal. To answer (c), we note that the
singularities for the four point function in Type II superstring amplitudes
should
 consist of branch cuts along the positive real axis, and simple and double
poles
 at even integers.
The appearance of any other type of singularity in addition to the above, such
as
poles or branch cuts or lacunae, would be inconsistent with unitarity,
causality
and Lorentz invariance.
We shall indeed see the physically correct
singularity structure emerge from the analytic continuation, essentially in
the following manner:
\medskip
\item{$$\bullet$$} the region $\tau_2 \rightarrow \infty$
produces branch cut singularities in
$s$, $t$, $u$, corresponding to two physical string intermediate states,
for example for the box diagram;
\item{$$\bullet$$} $z_i\sim z_j$ but all other $z$'s far away, leads to a
simple pole in the $s_{ij}$ channel;
\item{$$\bullet$$} $z_i$'s coming pairwise close together leads to double
poles.
\medskip
The singularities which arise from three $z_i$'s coming close together
correspond to
self-energy graphs for on-shell massless particles which vanish by space-time
supersymmetry.
The singularities with all four $z_i$'s close together correspond to one-loop
tadpole graphs, producing double poles in each channel.
For massless (dilaton) tadpoles, these contributions vanish to one-loop, but
there are also
scalar string states at higher mass levels which produce non-zero double pole
contributions with no branch cuts.

\bigskip
\centerline{\bf 3. One-loop Superstring Amplitudes and $\phi^3$ Box Diagrams}
\bigskip
We give now a detailed analysis of the integral representation (2.5). In view
of
 the
product formula for Jacobi $\vartheta$-functions, the integrand in (2.5)
involves an infinite number of factors. It is crucial to recognize which
factors can be expanded in
uniformly convergent series and treated perturbatively, and which ones cannot.
Mathematically, the resulting series need to be uniformly convergent.
Physically, the factors which cannot be expanded are the ones responsible for
the "stringy" aspects of the
theory, i.e., the ones not present in a mere infinite superposition of
quantum field theories.
\medskip
We begin by
decomposing the $z_i$ domain of integration into 6 regions according to the
various orderings of $\Im\, z_i$'s (c.f. [8][9][10]). The contributions are
the same for pairs of regions, and we get
$$
A(s,t,u)=2A(s,t)+2A(t,u)+2A(u,s)\eqno(3.1)
$$
where $A(s,t)$ is still defined by (2.5), but
the range of integration is now restricted to $$ \Im\, z_1\leq \Im\,
z_2\leq\Im\
,z_3\leq\Im\,z_4\eqno(3.2) $$ and $u$ has been set to $u=-s-t$. It is
convenient
to change variables to
$$
\eqalignno{
z_i-z_{i-1}&={\alpha_i\over 2\pi}+i\tau_2 u_i,\ i=1,\cdots,4\cr
z_4-z_0&=\tau \qquad q=e^{2\pi i \tau} &(3.3)\cr}
$$
Physically the parameters $\tau_2u_i$ correspond to evolution time between two
vertex operators
in the interaction picture of quantum mechanics, and the $\alpha_i$ angular
integrations to enforcing the constraint $L_0=\bar L_0$ on each string
propagator.
In terms of the
new variables $u_i$, $\alpha_i$, the amplitude $A(s,t)$ becomes
$$
A(s,t)=\int_F{d^2\tau\over\tau_2^2}\int_0^{2\pi}{d\alpha_i\over 2\pi}
\int_0^1du_i\delta(1-\sum_{i=1}^4u_i)|q|^{-su_1u_3-tu_2u_4}{\cal
R}(|q|^{u_i},\alpha_i;s,t)\eqno(3.4) $$
where the function ${\cal R}$ is given by an infinite product
$$
{\cal R}(|q|^{u_i},\alpha_i;s,t)
=\prod_{i\not=j}^4\prod_{n=0}^{\infty}|1-w_{ij}q^n|^{-s_{ij}}\eqno(3.5)
$$
Here $w_{ij}$ is defined as follows
$$
w_{ij}=\cases{e^{2\pi i(z_i-z_j)}, &Im$(z_i-z_j)>0$\cr
qe^{2\pi i(z_i-z_j)}, &Im$(z_i-z_j)<0$\cr}\eqno(3.6)
$$
for $i\not=j$. This definition has been arranged so that $|w_{ij}|\leq1$ and
$w_{ij}w_{ji}=q$
for any $i\not=j$. It is also convenient to identify the index $i=1$ with a
fifth index $i=5$. We have then $w_{(j+1)j}=|q|^{u_{j+1}}e^{i\alpha_{j+1}}$ for
$j=1,\cdots,4$. For future reference, we note also that $s_{54}=s_{14}=t$.
The integral representation (3.4) for
$A(s,t)$ shows that it is absolutely convergent for an entire strip in $s$ and
$t$
$$
\Re\, s,\Re\, t<0,\ \Re\, (s+t)>-2\eqno(3.7)
$$
Note, however, that all three expressions on the right hand side of
(3.1) are simultaneously convergent only when $\Re\,s=\Re\,t=\Re\,u=0$ when the
constraint
$s+t+u=0$ is enforced, in agreement with our earlier discussion. Thus we need
to
 analytically
continue $A(s,t)$ to the full cut plane $s,t\in{\bf C}\setminus{\bf R}_+$.
\medskip
Naively, we should expand the factor ${\cal R}$ into a series in $|q|^{u_i}$,
$i
=1\cdots,4$, {\it pointwise convergent} in the region $|w_{ij}|<1$
$$
{\cal R}(|q|^{u_i},\alpha;s,t)
=\sum_{n_i=0}^{\infty}\sum_{|\nu_i|\leq
n_i}P_{\{n_i\nu_i\}}^*(s,t)\prod_{i=1}^4
|q|^{u_in_i}e^{i\nu_i\alpha_i}\eqno(3.8)
$$
with the expansion coefficients $P_{\{n_i\nu_i\}}^*(s,t)$ polynomials in $s$
and
 $t$, and
construct an analytic continuation for expressions of the form (3.4) with
${\cal R}$ replaced by
a monomial of the form $\prod_{i=1}^4|q|^{n_iu_i}e^{i\nu_i\alpha_i}$. However,
this would not produce the intermediate poles on top of cuts peculiar to string
theory, as the following comparison with $\phi^3$ field theory illustrates.
\bigskip
First consider the contribution of a single monomial
term in the expression (3.8) inserted into the integral representation
(3.4). We shall see later that we may, without affecting the poles on top of
cuts, truncate the
fundamental domain $F$ to the simpler region $\{\tau_2\geq1,
|\tau_1|\leq1/2\}$.
 Thus $\tau_1$
becomes an angular variable, and the four angular integrations in (3.4) pick
out
 the terms with
$\nu_i=0$ for $i=1,\cdots,4$ in (3.8).
The remaining integral is of the form
$$
\int_1^{\infty}{d\tau_2\over\tau_2^2}\int_0^1du_i\delta(1-\sum_{i=1}^4u_i)
e^{2\pi\tau_2(su_1u_3+tu_2u_4-\sum_{i=1}^4n_iu_i)}
\eqno(3.9)$$
where the integers $n_i$ are actually positive and even.

Consider now a $\phi^3$-like box diagram in $d$
space-time dimensional quantum field theory with arbitrary masses $m_i$ for
each
 of the
propagators, and massless external on-shell states. Couplings are assumed to be
identical
for any choice of $m_i$ and are non-derivative $\phi^3$. The box diagram
(Euclidean) Feynman integral
can be performed as usual after introducing Feynman parameters $u_i$ and
exponentiating the denominator. We obtain
$$
\eqalignno{\epsfysize1.5in\vcenter{\epsfbox{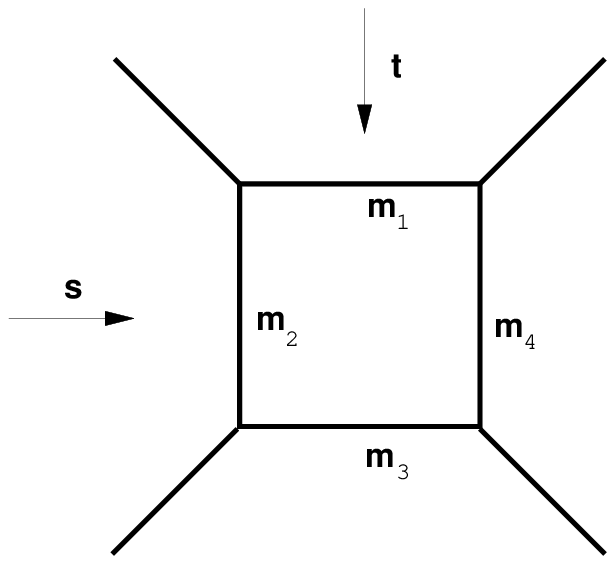}}
= &\int d^dk\prod_{i=1}^4{1\over (k+p_i)^2+m_i^2} \cr
=& \int_0^{\infty}{d\tau\over\tau^{{d\over 2}-3}}\int_0^1du_i
\delta(1-\sum_{i=1}^4u_i)
e^{2\pi\tau(su_1u_3+tu_2u_4-\sum_{i=1}^4u_im_i^2)} \cr}
$$
The $\phi^3$-like box diagram is almost identical to the partial superstring
amplitude provided the following identifications are made. Clearly the
space-time dimension should be
$d=10$, as expected, and the masses $m_i^2$ should be positive even integers,
consistent with the superstring spectrum. A well-understood difference is that
the $d\tau_2$ integral in the superstring
amplitude is truncated away from 0. This is a remnant of duality, which results
in modular invariance and restricts the moduli integration to a fundamental
domain
 for $SL(2,{\bf Z})$,
thus making the theory finite in the ultraviolet. The more important point is
that
 there must
be subtleties in the resummation of the monomial contributions (3.9), since
otherwise we wouldhave no poles. These subtleties are manifestations of the
differences between string theory and an infinite superposition of field
theories.
\bigskip
\centerline{\bf 4. Single and Double Dispersion Relations}
\bigskip
\noindent{\it Double Dispersion Relations}

The analytic continuation of the amplitude $A(s,t)$ is based on a double
dispersion
relation for each term in the following expansion of the factor
 ${\cal R}(|q|^{u_i},\alpha_i;s,t)$ $$
{\cal R}(|q|^{u_i},\alpha_i;s,t)
=\prod_{i=1}^4 \bigl |1-e^{i\alpha_i}|q|^{u_i} \bigr |^{-s_i}\sum_{n_i=0}^{
\infty}\sum_{|\nu_i|\leq
n_i}P_{\{n_i\nu_i\}}(s,t)\prod_{i=1}^4|q|^{n_iu_i}e^{i\nu_i\alpha_i}
\eqno(4.1)
$$
where we have set $s_i=s$ for $i$ even, and $s_i=t$ for $i$ odd. The expansion
(4.1) is obtained from (3.5) by keeping the first four factors
$\prod_{i=2}^5|1-w_{i(i-1)}|^{-s_{i(i-1)}}$ and expanding all the remaining
ones. We shall see
that the first four factors are precisely the ones responsible for stringy
effects,
 and in
particular poles on top of cuts. As in (3.6), the coefficients
$P_{\{n_i\nu_i\}}
(s,t)$ are polynomials in $s$ and $t$. They can be obtained explicitly by
simple
recursive formulas, but
we shall not need these here. The desired analytic continuation of $A(s,t)$ to
an arbitrary half-space $\Re\,s,\Re\,t<N$ can now be obtained as follows.
\medskip
\noindent{\bf Lemma 1}. {\it For any positive integer} $N$, {\it we can write}
$$
A(s,t)
=\sum_{n_1+\cdots+n_4\leq 4N}\sum_{|\nu_i|\leq
n_i}P_{\{n_i\nu_i\}}(s,t)A_{\{n_i
\nu_i\}}(s,t)
+M_N(s,t)\eqno(4.2)
$$
{\it where} $M_N(s,t)$ {\it is a meromorphic function in the region}
$\Re\,s,\Re
\,t<N$,
{\it and the
amplitudes} $A_{\{n_i\nu_i\}}(s,t)$ {\it are defined by}
$$
\eqalignno{A_{\{n_i\nu_i\}}(s,t)
=\int_1^{\infty}{d\tau_2\over\tau_2^2}&\int_0^{2\pi}{d\alpha_i\over
2\pi}\int_0^
{1}
du_i\delta(1-\sum_{i=1}^4u_i)|q|^{-su_1u_3-tu_2u_4}\cr
&\times\prod_{i=1}^4 \bigl |1-e^{i\alpha_i}|q|^{u_i}\bigr |^{-s_i}|q|^{n_iu_i}
e^{i\nu_i\alpha_i}&(4.3)\cr}
$$
\medskip
The lemma is established by isolating from the integral (3.4) contributions
which can be given an independent meromorphic continuation in the region
$\Re\,s
,\Re\,t<N$.
Particular care is required to treat the eight factors in (3.5) corresponding
to
$n=0$ which have been expanded into series. Details can be found in [5]. We
turn
 to the analytic continuation of $A_{\{n_i\nu_i\}}(s,t)$.
The angular integrations are decoupled now, and may be evaluated in terms of
Gauss' hypergeometric functions $F(a,b;c;x)$
$$
\int_0^{2\pi}{d\alpha\over 2\pi}e^{i\alpha\nu}|1-xe^{i\alpha}|^{-s}=
C_{|\nu|}(s)x^{|\nu|}F({s\over 2},{s\over 2}+|\nu|;|\nu|+1;x^2)
$$
Here we have denoted by $C_n(s)$ the following expression closely related to
the
 tree-level
amplitude of an intermediate state of mass $2n$ with two external massless
state
s
$$
C_n(s)={\Gamma({s\over 2}+n)\over\Gamma({s\over 2})\Gamma(n+1)}
$$
Thus the amplitude $A_{\{n_i\nu_i\}}(s,t)$ becomes
$$
\eqalignno{A_{\{n_i\nu_i\}}(s,t)
=\int_1^{\infty}{d\tau_2\over\tau_2^2}\int_0^1
du_i&\delta(1-\sum_{i=1}^4u_i)|q|^{-su_1u_3-tu_2u_4+
\sum_{i=1}^4u_i(n_i+|\nu_i|)}\cr
&\times
\prod_{i=1}^4C_{|\nu_i|}(s_i)F({s_i\over 2},{s_i\over
2}+|\nu_i|;|\nu_i|+1;|q|^{2u_i})&(4.4)\cr}
$$
The main
ingredients needed for the analytic continuation of (4.4) are the Mellin and
the
 inverse
Laplace transforms of the hypergeometric functions. The former is defined by
$$
f_{n\nu}(s;\alpha)=C_{|\nu|}(s)\int_0^1dx\, x^{-1-\alpha+n+|\nu|}F({s\over 2},
{s\over2}+|\nu|;|\nu|+1;x^2)
$$
and can be shown, using the reflection formula for hypergeometric functions, to
admit a
meromorphic extension in both $s$ and $\alpha$ throughout the full complex
plane
, with simple
poles in $\alpha$ at evenly spaced integers starting from $-n-|\nu|$ and
in $s$ at positive integers. The latter is defined by
$$
C_{|\nu|}(s)x^{n+|\nu|}F({s\over 2},{s\over
2}+|\nu|;|\nu|+1;x^2)=\int_0^{\infty}d\beta\, x^{\beta}\varphi_{n\nu}(s;\beta)
$$
We set $\varphi_{n\nu}(s;\beta)=0$ for $\beta<0$. The Mellin and the inverse
Laplace transforms are related by
$$
\eqalign{
\varphi_{n\nu}(s,\beta)&={1\over 2\pi i}(f_{n\nu}(s;\beta+i\epsilon)
-f_{n\nu}(s;\beta-i\epsilon))\cr
f_{n\nu}(s;\alpha)&=\int_0^{\infty}d\beta~{\varphi_{n\nu}
(s;\beta)\over\beta-\alpha}\cr}
$$
Since $f_{n\nu}(s;\beta)$ is a meromorphic function of $s$, its discontinuity
is
 also
meromorphic in $s$, and in fact one can easily see that
$\varphi_{n\nu}(s;\beta)$
 is an entire function of $s$. Finally we set
$$
\Psi_{n_i\nu_i}(s,t;\beta_i)=\prod_{i=1}^4\varphi_{n_i\nu_i}(s_i;\beta_i)
$$
\bigskip

\noindent{\bf Theorem 1}. {\it The amplitudes}
$A_{\{n_i\nu_i\}}(s,t)$ {\it can be expressed as}
$$
A_{\{n_i\nu_i\}}(s,t)=\int_0^{\infty}\int_0^{\infty}{\rho_{\{n_i\nu_i\}}
(s,t;\sigma,\tau)\over(s-\sigma) (t-\tau)}d\tau d\sigma
+ M_{\{n_i\nu_i\}} (s,t) \eqno(4.5)
$$
{\it where the density} $\rho_{\{n_i\nu_i\}}$ {\it is given by}
$$
\eqalignno{
\rho_{\{n_i\nu_i\}}(s,t;\sigma,\tau) =& \int_0^\infty d\beta_1
\int_0^\infty d\beta_2
\int_0^1 du_1
\int_0^{1-u_1} du_2
(1-u_1-u_2)^2
\int_{x_0}^\infty dx (x-x_0)^2 \cr
&
\quad \times \Psi_{n_i\nu_i}(s,t;\beta_1,\beta_2, u_1 \sigma -x,
u_2 \tau -x)
&(4.6)\cr}
$$
{\it Here} $x_0$ {\it denotes} $(u_1\beta_1+u_2\beta_2)(1-u_1-u_2)^{-1}$
{\it and}
$M_{\{n_i\nu_i\}} (s,t)$ {\it is a globally meromorphic function of} $s$
{\it and} $t$. {\it The
integral (4.5) defines (after a subtraction of a meromorphic function) a
function of} $s,t$ {\it meromorphic in the cut plane} $s,t\in{\bf C}
\setminus{\bf R}_+$. {\it More precisely, the
domain of holomorphy of} $A_{\{n_i\nu_i\}}(s,t)$ {\it is given by}
$$
\eqalign{
s&\in{\bf
C}\setminus\big[(\sqrt{n_1+|\nu_1|}+\sqrt{n_3+|\nu_3|})^2,+\infty\big]
\cr
t&\in{\bf
C}\setminus\big[(\sqrt{n_2+|\nu_2|}+\sqrt{n_4+|\nu_4|})^2,+\infty\big]
\cr}
$$

Together with Lemma 1, Theorem 1 provides a complete description of all the
 branch cut singularities
in the analytic continuation of the amplitude $A(s,t,u)$. Representations of
the
 form (4.5) are usually referred to as {\it double dispersion relations}.
 They were proposed by
Mandelstam and constituted a fundamental tool of particle physics in the 1950's
[11]. They are
analogues of the K$\ddot{\rm a}$llen-Lehmann representation we discussed
earlier
for the two-point function. The function $\rho_{\{n_i\nu_i\}}(s,t;\sigma,\tau)$
is referred to as
a {\it double spectral density} and is an entire function of $s$ and $t$ for
fixed $\sigma$ and $\tau$. To see this, we note that the range of integration
 in the expression (4.6) for
$\rho_{\{n_i\nu_i\}}$ is actually finite. Indeed $\varphi_{n\nu}(s;\beta)$
vanishes when $\beta<0$, and thus the regions contributing are
$$
u_1\sigma-x>0, u_2\tau-x>0
$$
Thus the $x$ range is bounded, and since $x>x_0>
(u_1\beta_1+u_2\beta_2)$, it follows also that $\beta_1<\sigma$ and
$\beta_2<\tau$. The domain of analyticity is a familiar one, as can be seen by
 introducing the mass value of the lowest branch cut : $n_i+|\nu_i|=m_i^2$, so
that the region becomes $s\in {\bf C}\setminus [(m_1+m_3)^2,\infty]$; $t\in
{\bf C}
\setminus [(m_2+m_4)^2,\infty]$.

\bigskip
\noindent{\it Single Dispersion Relations and Emergence of Poles}

In earlier discussions, we had emphasized that string amplitudes have
1-particle reducible contributions as well. The resulting poles occur both in
$M_{\{n_i\nu_i\}}$
and in the double spectral representation part. The latter ones are of
particular
 importance since they lead to decay
rates of massive strings, and can be recovered as follows.

Say we want to exhibit the poles in $s$ on top of the cut. We regard the double
dispersion
(4.5) as a single dispersion relation
$$
A_{\{n_i\nu_i\}}(s,t)=\int_0^{\infty}d\sigma{R_{\{n_i\nu_i\}}(s,t;\sigma)\over
s-\sigma}\eqno(4.7)
$$
with
simple spectral density
$$
R_{\{n_i\nu_i\}}(s,t;\sigma)=\int_0^{\infty}dt{\rho_{\{n_i\nu_i\}}(s,t;\sigma,
\tau)\over \tau-t}
$$
To understand its dependence on $s$ and $t$, we rewrite it as
$$
\eqalign{R_{\{n_i\nu_i\}}(s,t;\sigma)=\int_0^{\infty}&d\beta_1
\varphi_{n_1\nu_1}(t;\beta_1)
\int_0^{\infty}d\beta_2\varphi_{n_2\nu_2}(s;\beta_2)\int_0^1
du_1\int_0^{1-u_1}du_2\cr
&\times(1-u_1-u_2)^2\int_{x_0}^{\infty}(x-x_0)^2
\varphi_{n_3\nu_3}(t;u_1\sigma-x)f_{n_4\nu_4}(s;u_2t-x)\cr}
$$
This shows that $R_{\{n_i\nu_i\}}(s,t;\sigma)$ is an entire function of $t$.
This $t$-analyticity
may be used to expand $f_{n_4\nu_4}$ in terms of $t$, giving
$$
R_{\{n_i\nu_i\}}(s,t;\sigma)=\sum_{k_1,k_3=0}^{\infty}
C_{k_1}(t)C_{k_1+|\nu_1|}(t)C_{k_3}(t)C_{k_3+|\nu_3|}(t)
R_{\{n_i\nu_i\},k_1k_3}(s,t;\sigma)
$$
where
$$
\eqalignno{
R_{\{n_i\nu_i\},k_1k_3}(s,t;\sigma)=
\sum_{p=0}^{\infty}{2t^p\over p!(p+3)!}\int_0^1&du\,\theta(\sigma-{m_1^2\over
1-
u}-{m_3^2\over
u})\cr
&\times(\sigma
u(1-u)-m_3^2(1-u)-m_1^2u)^{p+3}\cr
&\times f_{n_2\nu_2}^{(p)}(s;m_3^2-u\sigma)f_{n_4\nu_4}^{(p)}
(s;m_3^2-u\sigma)&(4.8)\cr} $$
with $m_i^2\equiv 2k_i+n_i+|\nu_i|$. This expression shows that
$R_{\{n_i\nu_i\}}(s,t;\sigma)$ is a meromorphic function of $s$, with possibly
simple and double poles at even positive integers.
\medskip

We may represent the situation pictorially as follows. Recall that at tree
level,
the amplitudes may be expanded in poles, say in the $s$ channel (see Fig. 1).
\vskip -.75cm
$$
\epsfysize1.5in\vcenter{\epsfbox{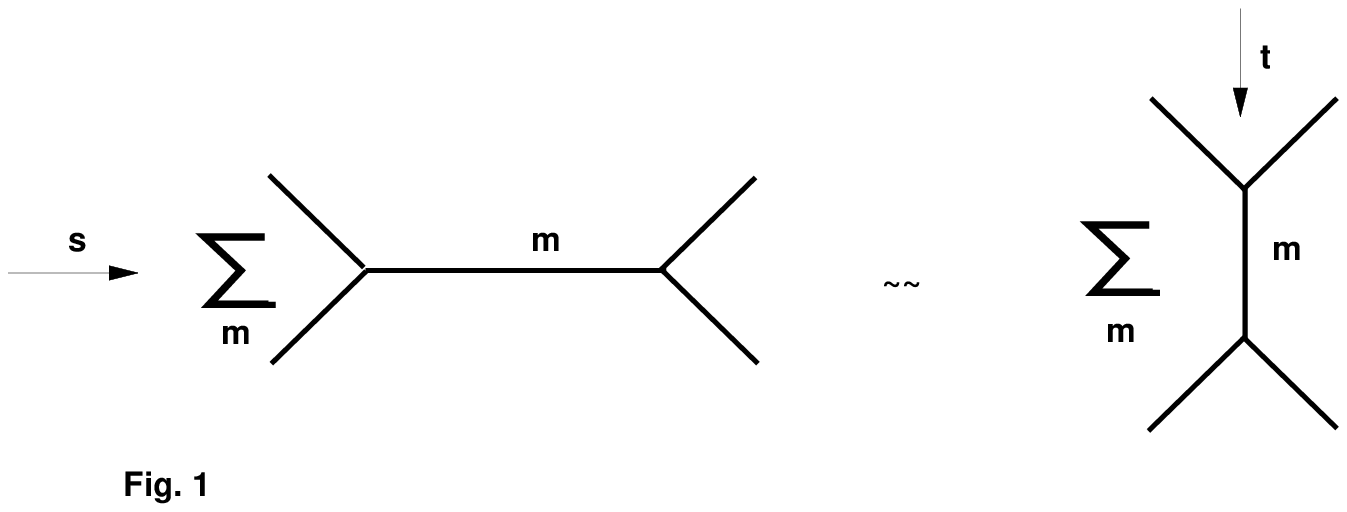}}
$$
But the series expansion on the right hand side of Fig. 1, which is equivalent
to a sum over field theory graphs in the $s$-channel, is not in general
 absolutely and uniformly convergent. As a result, the summed
series must be analytically continued, and in this analytically continued
answer, we find poles in the $t$ channel.
What is taking place here is the one-loop generalization of this phenomenon.
We had established previously a very close connection with a summation over all
masses of
$\phi^3$-like field theory box diagrams. But again this sum was not convergent,
and forced us
to treat the entire hypergeometric functions exactly. The analytically
continued
 sum as a
result exhibits pole singularities as well, as shown in Fig. 2 below.
$$
\epsfysize1.5in\vcenter{\epsfbox{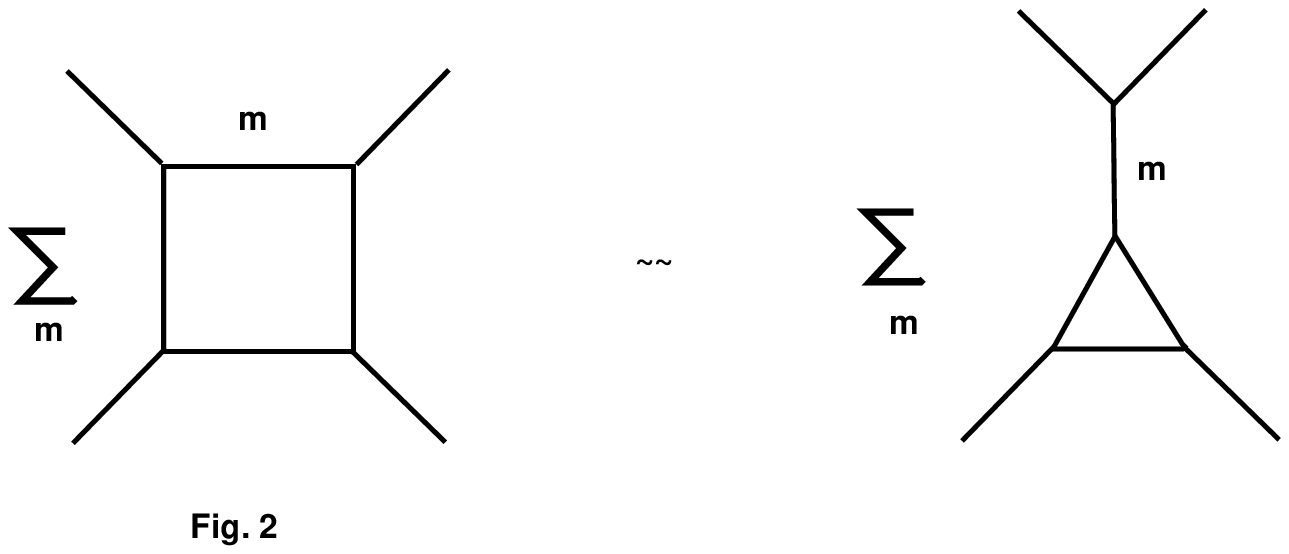}}
$$
\bigskip

\centerline{\bf 5. The $i\epsilon$ Prescription and Decay Rates of
Massive Strings}
\bigskip
We sketch here only two of the most immediate applications.
\bigskip
\noindent{\it The $i\epsilon$ prescription}

A physical amplitude
for real values of $s$, $t$, and $u$ can now be obtained by introducing the
$i\epsilon$ prescription in a straightforward way
$$
A_{i\epsilon}(s,t)
=
\int_0^{\infty}d\sigma\int_0^{\infty}d\tau{\rho(s,t;\sigma,\tau)
\over(s-\sigma+i\epsilon)(t-\tau+i\epsilon)}+M(s+i\epsilon,t+i\epsilon)
$$
so that the full amplitude is defined by
$$
A(s,t,u)=2A_{i\epsilon}(s,t)+2A_{i\epsilon}(t,u)+2A_{i\epsilon}(u,s)
$$
We note that this expression is different from a naive extrapolation when
$s,t,u\rightarrow s+i\epsilon,t+i\epsilon,u+i\epsilon$, which leads to an
incorrect singularity structure.
\bigskip
\noindent{\it Decay Rates of Massive Strings}

A practical application is the
calculation of decay rates for massive string states into 2 body decays of
strings of lesser mass to lowest order. As already pointed out, this decay
rate is simply related to the
imaginary part of the residue (i.e. coefficient in the Laurent expansion) of
the
 4-point amplitude at a double pole, say in $s$.
This problem has been considered in the Veneziano dual model already long ago
[12]. The residual $t$ dependence may be resolved in terms of the spin values
of
the decaying string state. Both $M$ and the double dispersion relation produce
 double poles, but it is straightforward to see that the residues at the double
poles in $M$ are real, and do not contribute to the
decay rates. The residue at $s=2N$ of the Mellin transform $f_{n\nu}(s;\alpha)$
$$
\lim_{s\rightarrow 2N}(s-2N)f_{n\nu}(s;\alpha)
$$
can be evaluated to be $F_{\nu}(2N,\alpha-n)$, with
$$
F_{\nu}(2N,\alpha)=-{1\over 2}
{\Gamma(-{\nu\over 2}-{\alpha\over 2})\Gamma({\nu\over 2}-{\alpha\over 2})
\over\Gamma(N)^2\Gamma(-{\nu\over 2}-{\alpha\over 2}+1-N)\Gamma({\nu\over 2}
-{\alpha\over 2}+1-N)}\eqno(5.1)
$$
It is a polynomial of degree $2N-2$ in $\alpha$. Substituting in (4.8), we find
\bigskip
\noindent{\bf Theorem 2}.
{\it The decay rate}
$$
\Gamma(2N,t)g^2C_{2N}(t)
\equiv \lim_{s\rightarrow2N}2\pi(s-2N)^2\int_0^{\infty}d\tau
{\rho(s,t;2N,\tau)\over t-\tau}
$$
{\it is given by the expression}
$$
\eqalignno{\Gamma(2N,t)g^2C_{2N}(t)=&\sum_{n_2,n_4=0}^{\infty}
\sum_{|\nu_i|\leq n_i}P_{\{n_i\nu_i\}}^{(2)}(s,t)
\sum_{p=0}^{\infty}{2t^p\over p!(p+3)!}\cr
&\times\int_0^1du\,\theta(2N-{m_3^2\over u}-{m_1^2\over
1-u})
(2Nu(1-u)-(1-u)m_3^2-um_1^2)^{p+3}\cr
&\qquad\times
F_{\nu_2}^{(p)}(2N,m_3^2-2Nu-n_2)F_{\nu_4}^{(p)}(2N,m_3^2-2Nu-n_4)&(5.2)\cr}
$$
{\it where} $P_{n_i\nu_i} ^{(2)} (s,t)$ {\it are polynomials in} $s$ {\it and}
$
t$ {\it which can
be obtained recursively through the following relations}
$$
\eqalign{
\int_0^{2\pi}\int _0 ^{2\pi} {d\alpha _1 d \alpha _3 \over 4\pi ^2}
{\cal R}(|q|^{u_i},\alpha _i;s,t) =&
\bigl | 1 - e^{i\alpha _2} |q|^{u_2} \bigr |^{-s}
\bigl | 1 - e^{i\alpha _4} |q|^{u_4} \bigr |^{-s} \cr
&\times \sum _{n_i =0} ^\infty \sum _{|\nu _j|\leq n_j}
P^{(2)} _{\{n_i\}\nu_2\nu_4}(s,t) |q|^{n_iu_i}
e^{i \alpha _2\nu_2+i\alpha _4\nu _4} \cr}
$$
\medskip
Note that all sums  in (5.2) are finite, actually bounded by
$2N-2$, signalling the fact that for fixed $N$, the number of channels is
finite
{}.
\bigskip
We conclude with a few remarks on some alternative approaches for calculating
decay rates [13]. It would of course be natural to calculate decay rates by
 squaring the corresponding tree level amplitudes and summing over all final
 polarizations and masses. Two problems arise then:
first, we must normalize string states correctly, which becomes prohibitively
difficult for higher
spin states. Second, we must actually perform the sum over final polarizations
 which again
is highly non-trivial. In fact, even in quantum field theory, one normally
prefers to calculate such decay rates by immediately considering the loop
amplitude,
 and taking its
imaginary part. Another method is to construct decay rates directly as a
one-loop amplitude with two massive vertex operators inserted. The problem with
this approach is that both vertex operators must have on-shell momenta, which
are
 fixed, and thus the amplitude has no more free parameters. But this amplitude
is
as divergent as the one-loop four-point function that we have been considering,
with
the difference that there is no variable left to analytically continue in. It
is
 not possible to make sense directly out of such amplitudes, without using the
 result that an analytic continuation exists, as derived in this paper.
 \bigskip
\bigskip

\centerline{\bf Acknowledgements}
\bigskip
E. D. gratefully acknowledges the hospitality extended to him
by the Institute for Theoretical Physics at Santa Barbara,
by the Yukawa Institute for Theoretical Physics at Kyoto University and in
particular by Takeo Inami, Satoshi Matsuda and Hirosi Ooguri.
D. H. P. would like to thank the Institute for Theoretical and Experimental
Physics in Moscow,
the organizers of the Alushta conference, and especially Andrei Marshakov
and Alexei Morozov for the very warm hospitality extended to him during his
stay in Russia and Crimea.

\bigskip
\bigskip
\centerline{\bf References}
\bigskip
\item{[1]}  K. Aoki, E. D'Hoker, and D.H. Phong, Nucl. Phys. B 342 (1990) 149.
\item{[2]} M.B. Green and J. Schwarz, Phys. Lett. 109 B (1982) 444.
\item{[3]} M.B. Green, J. Schwarz, and E. Witten, {\it Superstring Theory},
Cambridge University Press, 1987.
\item{[4]} E. D'Hoker and D.H. Phong, Rev. Mod. Physics 60 (1988) 917.
\item{[5]} E. D'Hoker and D.H. Phong, Phys. Rev. Lett. 70 (1993) 3692.
\item{[6]} G. Chew, {\it The Analytic S-matrix}, W. A. Benjamin (1966).
\item{[7]} R.J. Eden, P.V. Landshoff, D.I.Olive and J.C. Polkinghorne, {\it The
	Analytic S-Matrix}, Cambridge University Press (1966).
\item{[8]} M.B. Green, J.H. Schwarz and L. Brink, Nucl. Phys. B198 (1982) 474.
\item{[9]} K. Amano, Nucl. Phys. B328 (1989) 510.
\item{[10]} J. Montag and W.I. Weisberger, Nucl. Phys. B 363 (1991) 527.
\item{[11]} S.Mandelstam, Phys. Rev. 112 (1958) 1344; 115 (1959) 1741.
\item{[12]} C. B. Chiu and S. Matsuda, Nucl. Phys. B 134 (1978) 463; V. A.
Miranski,
V. P. Shelest, B. V. Struminsky, and G. M. Zinovjev, Phys. Lett. B 43 (1973)
73.
\item{[13]} K. Amano and A. Tsuchiya, Phys. Rev. D 39 (1989) 565; B. Sundborg,
Nucl. Phys. B 328
(1989) 415; D. Mitchell, N. Turok, R. Wilkinson, and P. Jetzer, Nucl. Phys. B
31
5 (1989) 1;
N. Marcus, Phys. Lett. B 219 (1989) 265; A. Berera, Report No. LBL-32248
(1992).

\end